\documentclass[12pt,preprint]{aastex}
\usepackage{psfig}
\newcommand{\beq}{\begin{equation}}
\newcommand{\eeq}{\end{equation}}
\newcommand{\beqn}{\begin{eqnarray}}
\newcommand{\eeqn}{\end{eqnarray}}
\newcommand{\no}{\noindent}
\newcommand{\non}{\nonumber}
\begin{document}

\title{\bf Construction of Coupled
Period-Mass Functions in Extrasolar Planets through the
Nonparametric Approach }

\author{Ing-Guey Jiang$^a$, Li-Chin Yeh$^b$, Yen-Chang Chang$^b$,
 Wen-Liang Hung$^c$}

\affil{\small $^a$Department of Physics and Institute of Astronomy,\\
National Tsing Hua University, Hsin-Chu, Taiwan\\
$^b$Department of Applied Mathematics,\\
National Hsinchu University of Education, Hsin-Chu, Taiwan \\
$^c$Graduate Institute of Computer Science,\\
National Hsinchu University of Education, Hsin-Chu, Taiwan}

\begin{abstract}

Using the period and mass data of two hundred and seventy-nine
extrasolar planets, we have constructed a coupled period-mass
function through the non-parametric approach. This analytic
expression of the coupled period-mass function has been obtained
for the first time in this field. Moreover, due to a moderate
period-mass correlation, the shapes of mass/period functions vary
as a function of period/mass. These results of mass and period
functions give way to two important implications: (1) {\it the
deficit of massive close-in planets is confirmed}, and (2) {\it
the more massive planets have larger ranges of possible semi-major
axes}. These interesting statistical results will
provide important clues into the theories of planetary formation.

\end{abstract}


\section{Introduction and Motivation}

After the first detection of an extra-solar planet (exoplanet)
around a millisecond pulsar in 1992 (Wolszczan \& Frail 1992), it
was soon reported that another exoplanet,
the first one around a
sun-like star, i.e. 51 Pegasi b, was found (Mayor \& Queloz 1995).
Ever since then, there
has been a continuous flood of discoveries of extra-solar planets.
As of February 2008, more than 200 planets have been detected
around solar type stars. These discoveries have led to a new era
in the study of planetary systems. For example, the traditional
theory for the formation of the Solar System does not likely
explain certain structures of extra-solar planetary systems. This
is due to the properties, discovered in extra-solar planetary
systems, being quite unlike our own. Many detailed simulations and
mechanisms have been proposed to explore these important issues
(Jiang \& Ip 2001, Kinoshita \& Nakai 2001,
 Armitage et al. 2002,  Ji et al. 2003, Jiang \& Yeh 2004a, Jiang \& Yeh 2004b, Boss 2005,
  Jiang \& Yeh 2007, Rice et al. 2008).

As the number of detected exoplanets keeps increasing, the
statistical properties of exoplanets have become more meaningful.
For example, assuming that the mass and period distributions are
two independent power-law functions, Tabachnik \& Tremaine (2002)
used the maximum likelihood method to determine the best
power-index. However, the possibility of a mass-period correlation
is not addressed in their work. Zucker \& Mazeh (2002) determined
the correlation coefficient between mass and period in logarithmic
space and concluded that the mass-period correlation is
significant.

On the other hand, a clustering analysis of the data we have on
exoplanets also gives some interesting results. Jiang et al.
(2006) took a first step into clustering analysis and found that
the mass distribution is continuous, and the orbital population
could be classified into three clusters which correspond to the
exoplanets in the regimes of tidal, ongoing tidal and disc
interaction. Marchi (2007) also worked on clustering through
different methods.

To take things a step further from the mass-period distribution
function of Tabachnik \& Tremaine (2002) and the mass-period
correlation of Zucker \& Mazeh (2002), Jiang, Yeh, Chang, \& Hung
(2007) (hereafter JYCH07) employed an algorithm to construct a
coupled mass-period function numerically. They were able to
include the possible correlation of mass and period into the
distribution function for the first time in this field and
obtained a distribution function that found a correlation to be
consistent. In fact, the mass-period distribution obtained by
JYCH07 should be called the mass-period {\it probability density
function} (pdf) in statistics. The integral of pdf is then called
the {\it cumulative distribution function} (cdf). We will use the
above terms in this paper.

Although JYCH07 successfully constructed the coupled mass-period
pdf numerically, due to constraints in the algorithm they
employed, they were forced to use the parametric approach of
$\beta$-distribution on the pdf fitting. The pdf is a basic
characteristic describing the behavior of random variables, i.e.
mass and period, and is so important that one has to choose the
underlying functional form carefully. One possibility to address
this problem is to use the nonparametric approach. This is because
the nonparametric approach is a distribution-free inference. That
is, an inference that is made without any assumptions regarding
the functional form of the underlying distribution. In addition,
the most valuable indication of the nonparametric approach is to
let the data speak for itself. We therefore see no other
reasonable course of action than to use the nonparametric approach
in this paper.

Moreover, we still consider the period-mass coupling even while
the pdf and cdf are being constructed. In order to make it
possible to proceed, we will employ a method called ``Copula
Modelling'' to obtain the coupled pdf and cdf on the period and
mass of exoplanets. This method is more general than the one used
in JYCH07 so that a nonparametric approach can be used to obtain
the coupled pdf. ``Copula Modelling'' has a long history of
development and was too complicated to be used with real data, in
practical terms, until Trivedi \& Zimmer (2005) clearly
demonstrated a standard modelling procedure.

In \S 2, we briefly describe the data and in \S 3, an estimation
of the nonparametric approach will be done. In \S 4, we introduce
the method of Copula Modelling and demonstrate its credibility.
The Copula Modelling will then be directly applied on the data of
exoplanets. The results will be described and discussed in
 \S 5. Our main conclusions will be found in \S 6.

\section{The Data}

We took samples of exoplanets from The Extrasolar Planets
Encyclopaedia (http://\\
exoplanet.eu/catalog-all.php),  2008 April 10. Our samples do not
include OGLE235-MOA53b, 2M1207b, GQ Lupb, AB Pic b, SCR 1845b,
UScoCTIO108b, or SWEEPS-04 because either their mass or their
period data was not listed. The outlier, PSR B1620-26b, with a
huge period (100 years), is also excluded.

The data of orbital periods is taken directly from the table in
The Extrasolar Planets Encyclopaedia. As a result, only the values
of projected mass ($m\ {\rm sin}i$) are listed and only a small
fraction of exoplanets' inclination angles $i$ are known so we
decided to provide two models of planetary mass in this paper. For
the ``minimum-mass model'', we simply set  ${\rm sin} i=1$ for all
planetary systems in the data. For the ``guess-mass model'', an
inclination angle $i$ within the observational constraint is
assigned to a planetary system through a random process and the
mass is then determined accordingly. In this case, if the
inclination angle $i$ is given in The Extrasolar Planets
Encyclopaedia for a particular planet, we simply use its value. If
there is no mention of observational constraints, the angle $i$
will be randomly chosen between $0^{\circ}$ and $90^{\circ}$.
Please note that the unit of period is days, and the unit of mass
is Jupiter Mass $(M_J)$.

\section{The Nonparametric Approach}

Considering $n$ data points of extrasolar planets in period and
mass spaces, i.e. $(p_1,m_1)$, $(p_2,m_2),\cdots,(p_n,m_n)$;
$F_P(p)$ and $F_M(m)$, are the cdfs of period and mass, and
$f_P$($p$) and $f_M$($m$) are the pdfs of period and mass,
respectively. An estimate of the cdf, $F_P(p)$, at the point $p$
is the proportion of samples that are less than or equal to $p$

\begin{eqnarray}
\hat F_P(p)=\frac{1}{n+1}\sum_{j=1}^n I(p_i\le p) \label{eq:f_p}
\end{eqnarray}
where $I(\cdot)$ is the indicator function defined by
\[ I(p_i\le
p)=\left\{\begin{array} {l@{\quad}l}
 1, & \mbox{if}~~p_i\le p,\\
 0, & \mbox{if}~~p_i>p.
\end{array}\right.
\]
Similarly, the nonparametric estimate of the cdf, $F_M(m)$, at the
point $m$ is
\begin{eqnarray}
 \hat F_M(m)=\frac{1}{n+1}\sum_{j=1}^n I(m_i\le m).\label{eq:f_m}
\end{eqnarray}
The solid curves in Figure 1(a)-(b) are $\hat F_P(p)$ and the
minimum-mass model's $\hat F_M(m)$. The dotted curve in Figure
1(b) is the guess-mass model's $\hat F_M(m)$.

To obtain the analytic expressions for the pdfs $f_P(p)$ and
$f_M(m)$, we first plot the histograms in $p$ and $m$ spaces, as
shown in Figure 1(c)-(d). In these two histograms, we choose the
bandwidths $h_P$ and $h_M$ of $p$ and $m$ as follows (Silverman
1986, page 47):
$$h_P=0.9 A_P n^{-1/5},~h_M=0.9 A_M n^{-1/5},$$
where
$$A_P=\min\Big\{S_P, \frac{IQR_P}{1.34}\Big\},~
A_M=\min\Big\{S_M, \frac{IQR_M}{1.34}\Big\},$$
 $S_P$ $(S_M)$ and $IQR_P$ $(IQR_M)$ are the standard deviation and
 interquartile range of $p_1,\cdots,p_n$ $(m_1,\cdots m_n)$, respectively.
Here the interquartile range is the difference between the first
and third quartiles (also see this definition in \S 5.1). In our
data, $S_P=896.464$, $IQR_P=846.360$, and $S_M=3.499$
($S_M=5.235$), $IQR_M=2.520$ ($IQR_M=3.694$) for the minimum-mass
model (guess-mass model).

We then use the adaptive kernel method (Silverman 1986, page 101)
to estimate the pdfs $f_P(p)$ and $f_M(m)$ as follows:

\begin{itemize}
 \item[{\bf Step 1}] Finding the pilot estimates
$$\tilde f_P(p)=\frac{1}{nh_P}\sum_{j=1}^n  K\Big(\frac{p-p_j}{h_P}\Big),$$
$$\tilde f_M(m)=\frac{1}{nh_M}\sum_{j=1}^n  K\Big(\frac{m-m_j}{h_M}\Big),$$
  where $K(\cdot)$ is the Gaussian kernel, i.e.,
  $$K(x)=\frac{1}{\sqrt{2\pi}}e^{-x^2/2}.$$

 \item[{\bf Step 2}] Defining local bandwidth factors $\lambda_i^P$ and
 $\lambda_i^M$ by
 $$\lambda_i^P=\Big(\frac{\tilde f_P(p_i)}{g_P}\Big)^{-1/2},~
\lambda_i^M=\Big(\frac{\tilde f_M(m_i)}{g_M}\Big)^{-1/2},
i=1,\cdots,n,$$
 where $g_P$ ($g_M$) is the geometric mean of $\tilde f_P(p_i)$
 ($\tilde f_M(m_i)$), that is,
 $$\ln g_P=\frac{1}{n}\sum_{i=1}^n \ln \tilde f_P(p_i),~
   \ln g_M=\frac{1}{n}\sum_{i=1}^n \ln \tilde f_M(m_i).$$

 \item[{\bf Step 3}] Obtaining the adaptive kernel estimate $\hat f_P(p)$
($\hat f_M(m)$) of the pdf $f_P(p)$ ($f_M(m)$) by
 $$\hat f_P(p)=\frac{1}{n} \sum_{j=1}^n \frac{1}{\lambda_j^P h_P}
 K\Big(\frac{p-p_j}{\lambda_j^P h_P}\Big),$$
$$\hat f_M(m)=\frac{1}{n} \sum_{j=1}^n \frac{1}{\lambda_j^M h_M}
 K\Big(\frac{m-m_j}{\lambda_j^M h_M}\Big).$$
\end{itemize}

Thus, the analytic expressions are obtained. In order to compare
these with the histograms, we define $ \hat f^h_P(p)\equiv
Area_P\times\hat f_P(p)$, where $Area_P=51240.68$ is the area
under the histogram of the period in Figure 1(c). Similarly, we
set $\hat f^{hm}_M(m)\equiv Area_{Mm} \times \hat f_M(m)$, the
 minimum-mass model, where $Area_{Mm}=153.06$, and set
 $\hat f^{hg}_M(m)\equiv Area_{Mg} \times \hat f_M(m)$, the
guess-mass model, where $Area_{Mg}=217.20$. $\hat f^h_P(p)$ is
plotted as the solid curve in Figure 1(c), $\hat f^{hm}_M(m)$ is
the solid curve in Figure 1(d), and $\hat f^{hg}_M(m)$ is the
dotted curve in Figure 1(d).

\section{The Copula Modelling Method}

In this section, we will describe the procedure to construct a new
period-mass pdf, in which the possible period and mass correlation
is included. The Copula Modelling method, which is widely used to
construct multi-variate distributions (Genest \& MacKay 1986,
 Frees \& Valdez 1998, Klugman \& Parsa 1999 and Venter et al.2007),
 will be introduced in the first part of this section and
its credibility will be demonstrated in the second part.

\subsection{The Procedure and Equations}

Here we describe the Copula Modelling method in a way that readers
can reproduce the results or work on their own applications with
the equations provided in this paper. However, please refer to
Trivedi \& Zimmer (2005) for further details of Copula Modelling.

According to Trivedi \& Zimmer (2005), there are several ``copula
functions'' to be used but the Frank copula is more flexible as it
allows two variables of the data to have negative, zero, and
positive correlations (Frank 1979). This is suitable to our work,
as we hope that all different kinds of possible period-mass
correlations can be considered in the construction of coupled pdf.
The Frank copula function is given by
 \beq
 C(u_1,u_2;\theta)=\frac{-1}{\theta} \ln\Big[
  1+\frac{(e^{-\theta u_1}-1)(e^{-\theta u_2}-1)}{e^{-\theta}-1}\Big],
\eeq
 where $u_1$, $u_2$ ($0\le u_1, u_2\le 1$) are two marginal
distribution functions and $\theta$ ($-\infty<\theta<\infty$) is
the dependence parameter. Positive, zero and negative values of
$\theta$ correspond to the positive dependence, independence and
negative dependence between two marginal variables, respectively.

For our work here, $u_1$ is the cdf of period, $F_P$($p$), and
$u_2$ is the cdf of mass, $F_M(m)$. In Copula Modelling, the pdf
of the coupled period-mass distribution is
\begin{eqnarray}
 & & f_{(P,M)}(p,m|\theta) =\frac{\partial^2 C (F_P(p),F_M(m);\theta)}
 {\partial F_P \partial F_M} f_P(p) f_M(m), \non \\
&=&\frac{-\theta(e^{-\theta}-1)
 e^{-\theta F_P(p)}e^{-\theta F_M(m)}}
 {\Big[e^{-\theta}-1+(e^{-\theta F_P(p)}-1)(e^{-\theta F_M(m)}-1)\Big]^2}
 f_P(p) f_M(m).\label{eq:f_pm}
\end{eqnarray}

We now have an analytic form of the coupled period-mass pdf where
the parameter $\theta$ is to be determined through the Maximum
Likelihood Method.

 The log-likelihood function of $\theta$ for the samples $(p_i,m_i), i=1,..., n$
 can be written as

 \beq
\ell(\theta)=\ell_1+\ell_2(\theta),
 \label{eq:full_l}
 \eeq
  where
\begin{eqnarray}
 \ell_1 &=& \sum_{i=1}^n \Big[ \ln f_P(p_i)+\ln f_M(m_i)\Big]\\
\ell_2(\theta) &=& n\ln \big[-\theta~
 (e^{-\theta}-1)\big] -\sum_{i=1}^n \Big\{\theta
\big[F_P(p_i) + F_M(m_i) \big] \nonumber \\
 & & +2 \ln \big[e^{-\theta}-1+(e^{-\theta F_P(p_i)}-1)(e^{-\theta F_M(m_i)}-1)
\big]\Big\}.
\end{eqnarray}
Differentiating $\ell(\theta)$ with respect to $\theta$, we obtain
{\small
\begin{eqnarray*}
 \frac{\partial \ell(\theta)}{\partial \theta}=
 \frac{\partial \ell_2(\theta)}{\partial \theta}
 &=& \sum_{i=1}^n \Big\{\frac{e^{-\theta}-1-\theta e^{-\theta}}
 {\theta (e^{-\theta}-1)}-\big[F_M(m_i)+F_P(p_i)\big]\\
 & & + 2\cdot \frac{e^{-\theta}+F_M(m_i)e^{-\theta F_M(m_i)}
(e^{-\theta F_P(p_i)}-1)+
 F_P(p_i)e^{-\theta F_P(p_i)}(e^{-\theta F_M(m_i)}-1)}
 {e^{-\theta}-1+(e^{-\theta F_M(m_i)}-1)(e^{-\theta F_P(p_i)}-1)}\Big\}.
\end{eqnarray*}
}

 \normalsize
 After the estimates of cdfs $\hat F_P(p)$ and $\hat F_M(m)$ have been
 substituted into $\partial \ell(\theta)/\partial \theta$,
the estimate of $\theta$ is obtained by solving
$$ \frac{\partial \ell(\theta)}{\partial \theta}=0.$$

Moreover, according to Genets (1987), the parameter $\theta$ in
Copula Modelling is related to the Spearman rank-order correlation
coefficient ($\rho_S$) through the below formula:
\begin{eqnarray}
 \rho_S \approx \rho_{G}
\equiv (1-\theta e^{-\theta/2}-e^{-\theta})(e^{-\theta/2}-1)^{-2}.
\end{eqnarray}
We call this $\rho_{G}$ the Genets correlation coefficient in this
paper.

\subsection{The Credibility Test}

Since this is the first time that Copula Modelling has been
introduced and employed in astronomy, we shall demonstrate its
credibility. We will generate four sets of two hundred and
seventy-nine artificial data points of uniform random variables
$x$ and $y$, with different strength of $x$-$y$ correlations as
presented in Figure 2(a)-(d). The Spearman correlation
coefficients $\rho_S$ (also see \S 5.2 for the definition) between
$x$ and $y$ are in Table 1. We apply the Copula Modelling on these
four sets of experimental data, where the nonparametric approach
is used to obtain the cdfs of $x$ and $y$. Finally, the coupled
$x$-$y$ pdf, the coupling parameter $\theta$, and $\rho_G$ are
obtained. We also calculate the Bootstrap Confidence Interval
(C.I.) for $\theta$ and $\rho_G$ with the number of bootstrap
replications B=2000 (JYCH07). These results are all listed in
Table 1.

\vskip 0.2truein
 \begin{center}
 {\bf Table 1 }
\vskip 0.05truein
     \begin{tabular}{|c|c|c|l|c|l|}\hline
 Data Set & $\rho_S$  & $\theta$ & 95$\%$ C.I. for $\theta$ &
   $\rho_G$   &  95\% C.I. for $\rho_G$ \\ \hline
 (1) & 0.042  &  0.25 & (-0.455, 0.955) & 0.042 & (-0.076,0.158)\\\hline
 (2) & 0.231  & 1.42  & (0.685,2.140)   & 0.233 & (0.114,0.343)\\\hline
 (3) & 0.416  & 2.725 & (1.93,3.535)    & 0.427 & (0.312,0.534)\\ \hline
 (4) & 0.788  & 7.57  & (6.355,8.845)   & 0.866 & (0.798,0.915)\\ \hline
       \end{tabular}
\end{center}

Because the values of Spearman correlation coefficient $\rho_S$
are close to $\rho_G$ and within $\rho_G$'s 95\% confidence
intervals, we confirm that Copula Modelling gives the correct
coupling parameter $\theta$ and the Genets correlation
coefficients $\rho_G$. Thus, the coupling between $x$ and $y$ can
be correctly included when the pdf is constructed for any given
strength of correlation.

\section{Results}

In this section, the results of the coupled period-mass
distribution and the correlation coefficients will be presented.

\subsection{The Coupled Period-Mass Distribution}

Using the Copula Modelling, the estimate of $\theta$ is
$\hat\theta=2.3826$ for the minimum-mass model. Through the
bootstrap algorithm as described in JYCH07 with the number of
bootstrap replications $B=2000$, the standard error of
$\hat\theta$ is $0.3669$. In order to properly understand the
dependence parameter $\theta$, we also obtain the 95\% bootstrap
C.I. for $\theta$, which is $(1.6514, 3.1190)$. For the guess-mass
model, the estimate of $\theta$ is $\hat\theta=2.4565$ and its
95\% bootstrap C.I. is $(1.7282, 3.1633)$.

Furthermore, in order to check the stability of the guess-mass
model, we repeat the random process to generate 100 guess-mass
models and apply Copula Modelling on them. The average value of
$\hat\theta$ is $2.9249$ with the standard deviation $0.3349$. We
then employ the interquartile range (Turky 1977) to check for any
outliers of $\hat\theta$ from these 100 guess-mass models. The
interquartile range is the difference between the first quartile
$Q_1$ and the third quartile $Q_3$, i.e. $IQR=Q_3-Q_1$. Inner
fences are the left and right from the median at a distance of
$1.5$ times the $IQR$. Outer fences are at a distance of $3$ times
the $IQR$. The values lying between the inner and outer fences are
called suspected outliers and those lying beyond the outer fences
are called outliers (Hogg \& Tanis 2006).

The smallest, first quartile, median, third quartile and largest
of these 100 $\hat\theta$ values,
 denoted by $Min, Q_1, Me, Q_3, Max$, respectively, are
 $$Min=2.3730,~Q_1=2.6297,~Me=2.8833,~Q_3=3.1968,~Max=3.5776.$$
 Therefore, $IQR=0.5671$ and cutoffs for outliers are
$$Q_3+1.5 IQR=4.0475,~Q_3+3 IQR=4.8981,~Q_1-1.5 IQR=1.7791,~
 Q_1-3 IQR=0.9284.$$
Furthermore, we find that
$$Q_1-1.5 IQR(=1.7791)<Min(=2.3730)<Max(=3.5776)<Q_3+1.5IQR(=4.0475).$$
Thus, all 100 $\hat\theta$ values of the guess-mass model lie
within the inner fences. It means that no outliers exist in these
100 values and so the stability of the guess-mass model is
confirmed.

Figure 3 shows the three dimensional view of the coupled
period-mass pdf, $f_{(P,M)}(p,m|\theta)$, of the guess-mass model.
The contour of Figure 3 is presented in Figure 4. The plots of the
minimum-mass model's $f_{(P,M)}(p,m|\theta)$ are very similar to
the above, so we have not shown them.

We know that when the period and mass are completely independent,
$f_{(P,M)}(p,m|\theta)=f_P(p) f_M(m)$. Thus the term
$$
\frac{-\theta(e^{-\theta}-1) e^{-\theta F_P(p)}e^{-\theta F_M(m)}}
 {\left\{e^{-\theta}-1+(e^{-\theta F_P(p)}-1)(e^{-\theta F_M(m)}-1)\right\}^2}
$$
in Eq. (4) is the one to take the period-mass coupling into
account. We will hereafter call it the {\it Coupling Factor}. To
make it clear how the Coupling Factor behaves, its value as a
function of $p$ and $m$ of the guess-mass model is plotted in
Figure 5. Figure 6 is the color contour plot. It clearly shows
that the Coupling Factor becomes larger than one when both period
and mass are very small or when both of them are large (area of
red). It also shows that the Coupling Factor is less than one in
the blue area.

\subsection{The Correlation Coefficients}

JYCH07 calculated the linear correlation coefficients (also called
Pearson's correlation coefficients) in both $m$-$p$ and
 ${\rm ln} m$-${\rm ln} p$ spaces and found a weak correlation in $m$-$p$ and a
moderate correlation in ${\rm ln} m$-${\rm ln} p$ space. In order
to maintain a consistent determination on the correlation
coefficients, we now calculate the Spearman rank-order correlation
coefficients (Press et al. 1992), which are invariant under
 strictly increasing nonlinear transformations (Schweizer and Sklar 2005).

For pairs of quantities $(x_i, y_i), i=1,\ldots,n$, the linear
correlation coefficient is given by
\begin{eqnarray}
r=\frac{\sum_{i=1}^n (x_i-\bar x)(y_i-\bar y)}
 {\sqrt{\sum_{i=1}^n (x_i-\bar x)^2 \sum_{i=1}^n (y_i-\bar y)^2}},
\end{eqnarray}
 where $\bar x=\sum_{i=1}^n x_i/n$,  $\bar y=\sum_{i=1}^n y_i/n$.
The Spearman rank-order correlation coefficient is calculated by
the above formula with $x_i$ and $y_i$ replaced by their ranks.
Given $R(x_i)$ the rank of $x_i$ and $R(y_i)$ the rank of $y_i$,
then the Spearman rank-order correlation coefficient can be
written as
\begin{eqnarray}
\rho_S=1-\frac{6}{n^3-n}\sum_{i=1}^n \big[R(x_i)-R(y_i)\big]^2.
\end{eqnarray}
We note that $|\rho_S|=1$ indicates a perfect dependence and
$\rho_S=0$ means no dependence. When $\rho_S=1$ there is a direct
perfect dependence and when $\rho_S=-1$ there is an inverse
perfect dependence. Furthermore, according to Cohen (1988),
 $0.1<|\rho_S| \leq 0.3$  means the correlation is weak,
$0.3< |\rho_S| \leq 0.5$ indicates a moderate correlation, and
$0.5< |\rho_S| \leq 1.0$ is indicative of strong correlation.

For the minimum-mass model, the Spearman rank-order correlation
coefficient is obtained as $\rho_S=0.3769$. Through Copula
Modelling, we also find the estimate of $\rho_G$, which is
$\hat\rho_G=0.3792$. It is obvious that the Spearman rank-order
correlation coefficient $\rho_S=0.3769$ is very close to
$\hat\rho_G$. Moreover, the 95\% bootstrap C.I.  with the number
 of bootstrap replications $B=2000$ for $\rho_G$ is $(0.2691,0.4811)$.
For the guess-mass model, we have $\hat\rho_G=0.3899$ with a 95\%
bootstrap C.I. $(0.2811, 0.4869)$. These results are all
consistent and confirm that {\it there is a positive period-mass
correlation for exoplanets}.

\section{Conclusions}


Using the data of exoplanets, for the first time in this field we
have constructed an analytic coupled period-mass function through
a nonparametric approach. Moreover, we calculate the Spearman
rank-order correlation coefficient, which gives the same results
for linear and logarithmic spaces, and the results in the previous
section show that there is a  moderate positive period-mass
correlation.

In order to comprehend the implication of our results, in Figure
 7(a)-(b), we plot  $f_{(P,M)}(p,m|\theta)$ with $m=1, 5, 10, 15 M_J$
(i.e. the period functions given different masses), and also
$f_{(P,M)}(p,m|\theta)$ with $p=1, 50, 100, 150$ days (i.e. the
mass functions given different periods) in logarithmic spaces. For
purposes of comparing, $f_P(p)\times f_M(m)$ with $m=1, 5, 10, 15
M_J$ (the independent period functions)  and $f_P(p)\times f_M(m)$
with $p=1, 50, 100, 150$ days (the independent mass functions) are
also plotted in Figure 7(c)-(d). Of course, the shapes of
independent period functions with $m=1, 5, 10, 15 M_J$ are all the
same, and the shapes of independent mass functions given different
periods are all exactly the same as well.

We find that the period function of $m=1 M_J$ is very similar with
the independent period functions. However, the period functions of
$m=5, 10, 15 M_J$ are different from the independent ones, in a
way that the functions are lower at the smaller $p$ end and
slightly higher at the larger $p$ end.
Thus, the overall period functions of massive planets (say $m=5,
10, 15 M_J$) at large $p$ and small $p$ ends are closer than the
one of lighter planets (say $m=1 M_J$). Therefore,
the fractions of larger and smaller $p$ (or semi-major-axis)
planets are closer for those planets with mass $m=5, 10, 15 M_J$.

This implies that {\it the more massive planets have larger ranges
of possible semi-major axes}. This interesting statistical result
will provide important clues into the theories of planetary
formation.

On the other hand, the mass functions of $p=50, 100, 150$ days are
all very similar with the independent mass functions. However, the
mass function of $p=1$ day is different from the independent one
in a way that the function is higher at the smaller $m$ end and
lower at the larger $m$ end. Thus, the mass function of short
period planets (say $p=1$ day) is steeper than the one of long
period planets (say $p=50, 100, 150$ days). This implies that the
percentage of massive planets are relatively small for the short
period planets. This result reconfirms {\it the deficit of massive
close-in planets} due to tidal interaction as studied in Jiang et
al. (2003).


\section*{Acknowledgments}

We are grateful to the referee's suggestions. This work is
supported in part by the National Science Council, Taiwan.

\clearpage
\begin{figure}
\epsscale{1.0}
 \plotone{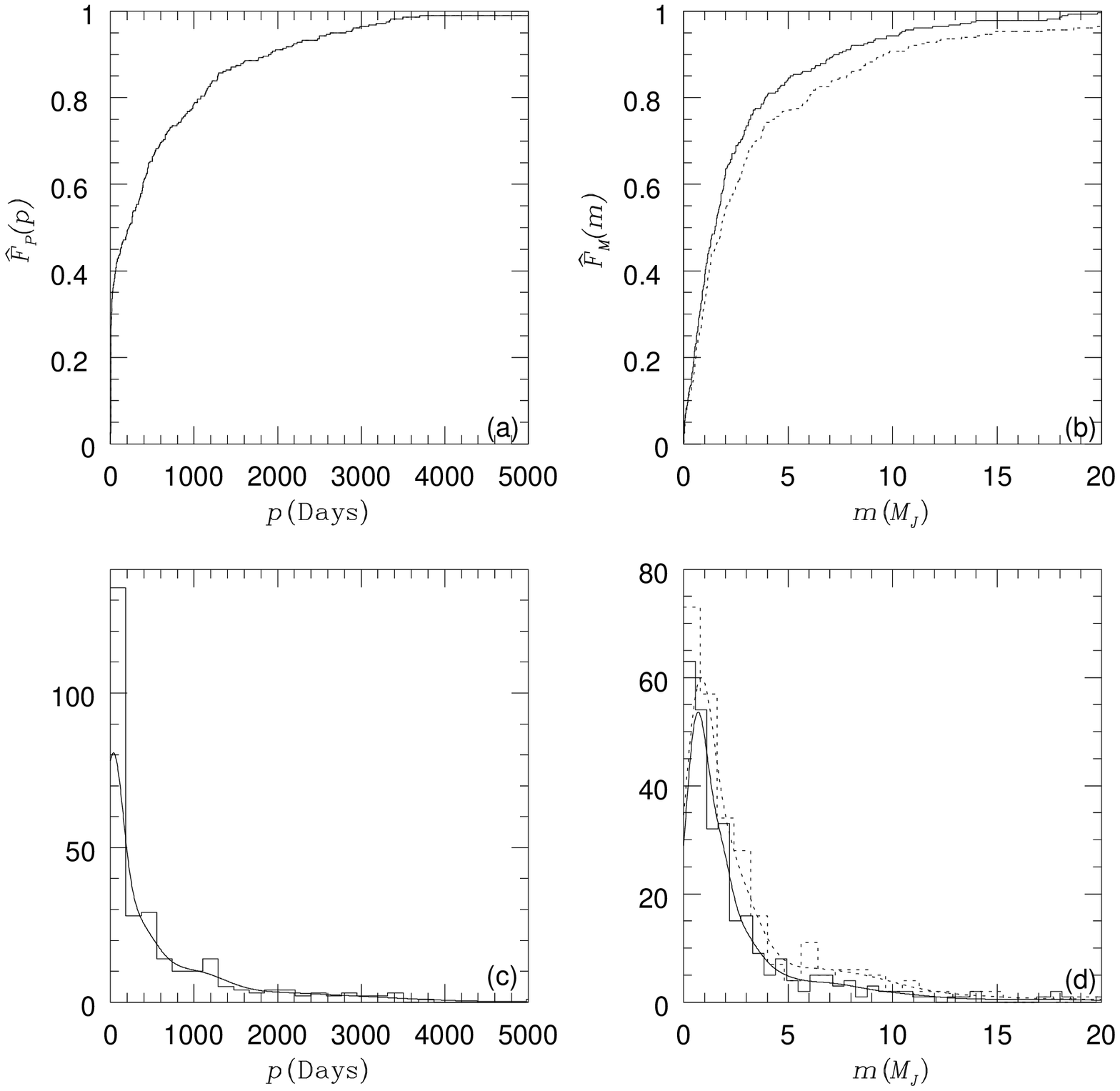}
 \caption{The cumulative distribution function (cdf)
and probability density function (pdf) of planetary period and
mass. (a) The period cdf. (b) The mass cdf of the minimum-mass
model (solid curve) and the guess-mass model (dotted curve). (c)
The histogram of planets in $p$ space and also the period pdf
$\hat f^h_P(p)$ (solid curve). (d) The histogram of planets in $m$
space of the minimum-mass model (solid line) and the guess-mass
model (dotted line), and also the mass pdf of the minimum-mass
model $\hat f^{hm}_M(m)$ (solid curve) and the guess-mass model
$\hat f^{hg}_M(m)$ (dotted curve).}

\end{figure}

\clearpage
\begin{figure}
\epsscale{1.0}
 \plotone{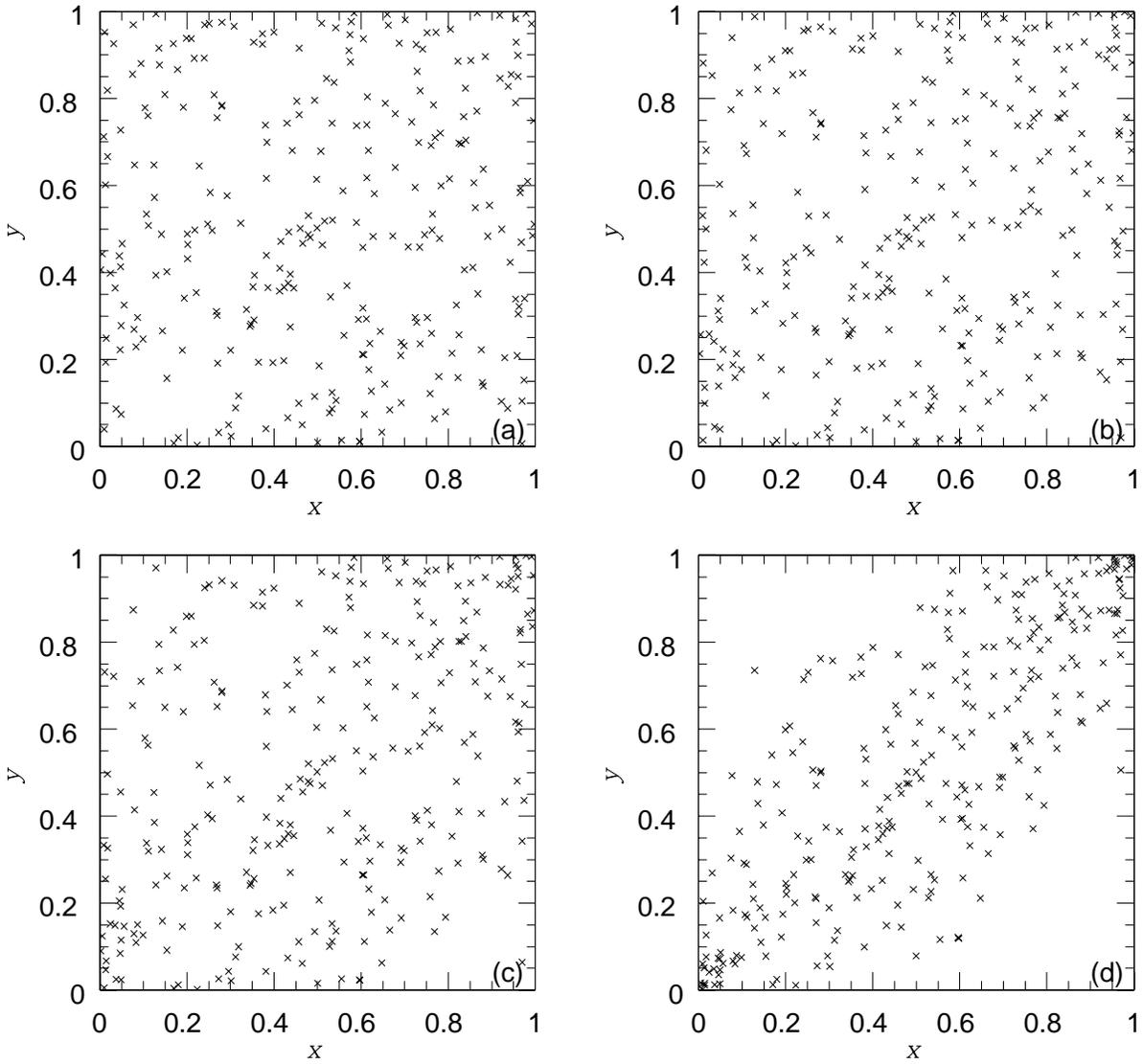}
 \caption{The random variables $x$ and $y$ in the credibility test of
Copula Modelling. (a) $\rho_S$= 0.042. (b) $\rho_S$=0.231. (c)
$\rho_S$= 0.416   (d) $\rho_S$= 0.788.}
\end{figure}

\clearpage
\begin{figure}
\epsscale{1.0}
 \plotone{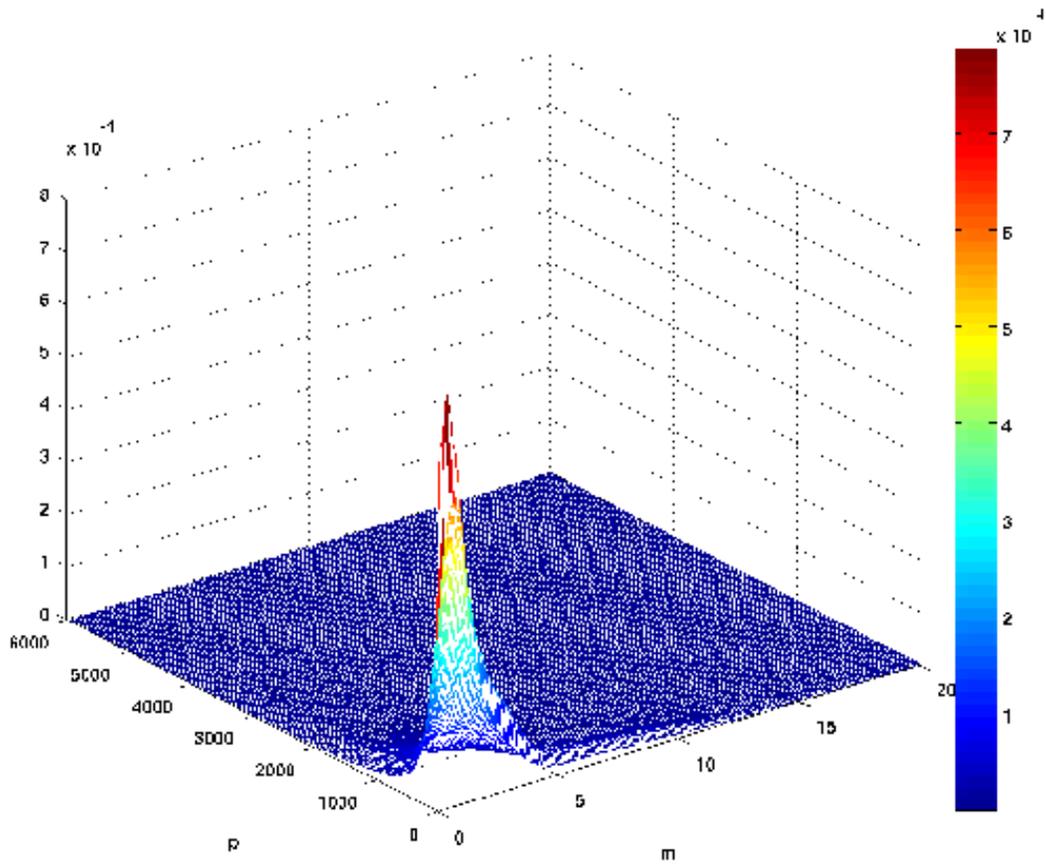}
 \caption{The three dimensional view of the coupled period-mass
pdf, $f_{(P,M)}(p,m|\theta)$, of the guess-mass model.}
\end{figure}

\clearpage
\begin{figure}
\epsscale{1.0}
\plotone{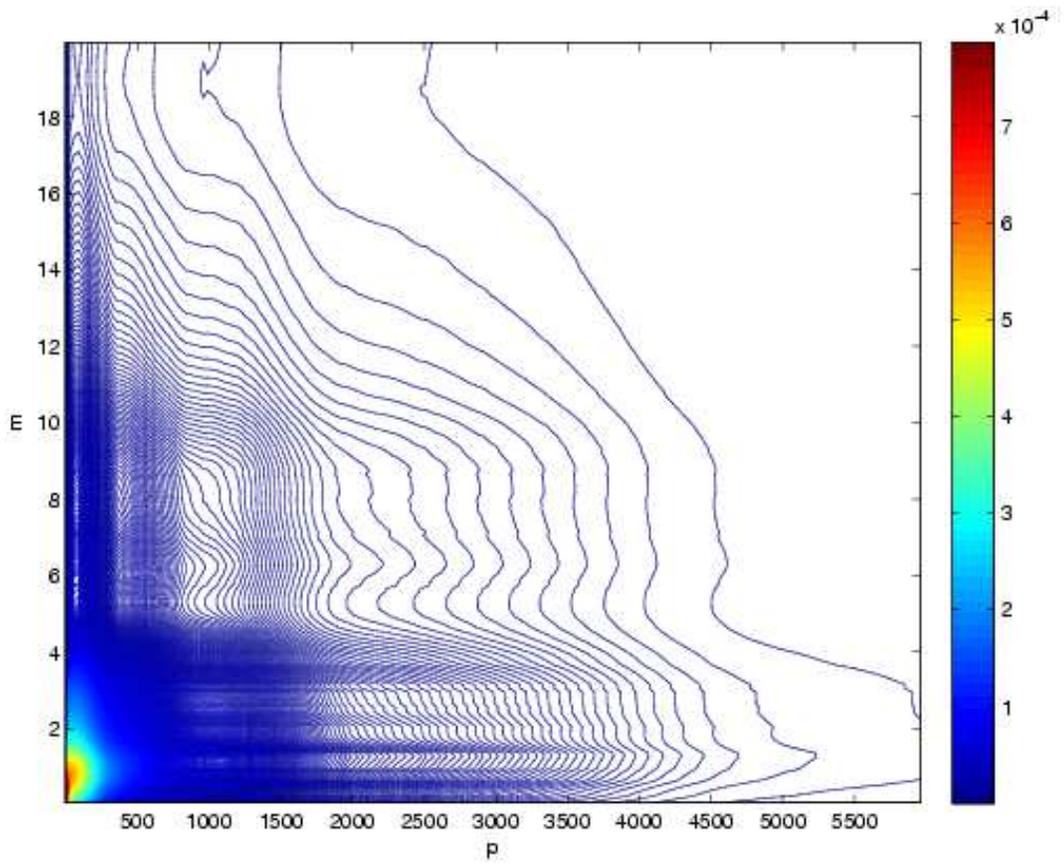}
\caption{The color contour of
the coupled period-mass pdf, $f_{(P,M)}(p,m|\theta)$, of the
guess-mass model.}
\end{figure}

\clearpage
\begin{figure}
\epsscale{1.0}
 \plotone{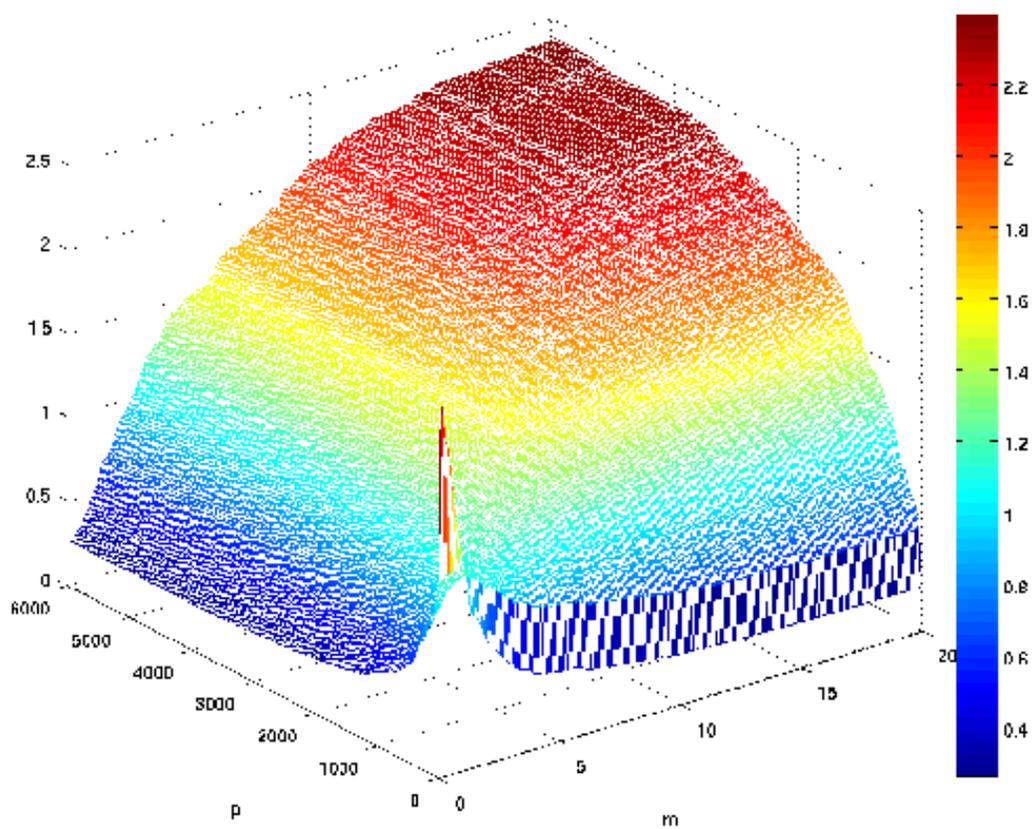}
 \caption{The three dimensional
view of the Coupling Factor of the guess-mass model.
}
\end{figure}

\clearpage
\begin{figure}
\epsscale{1.0}
 \plotone{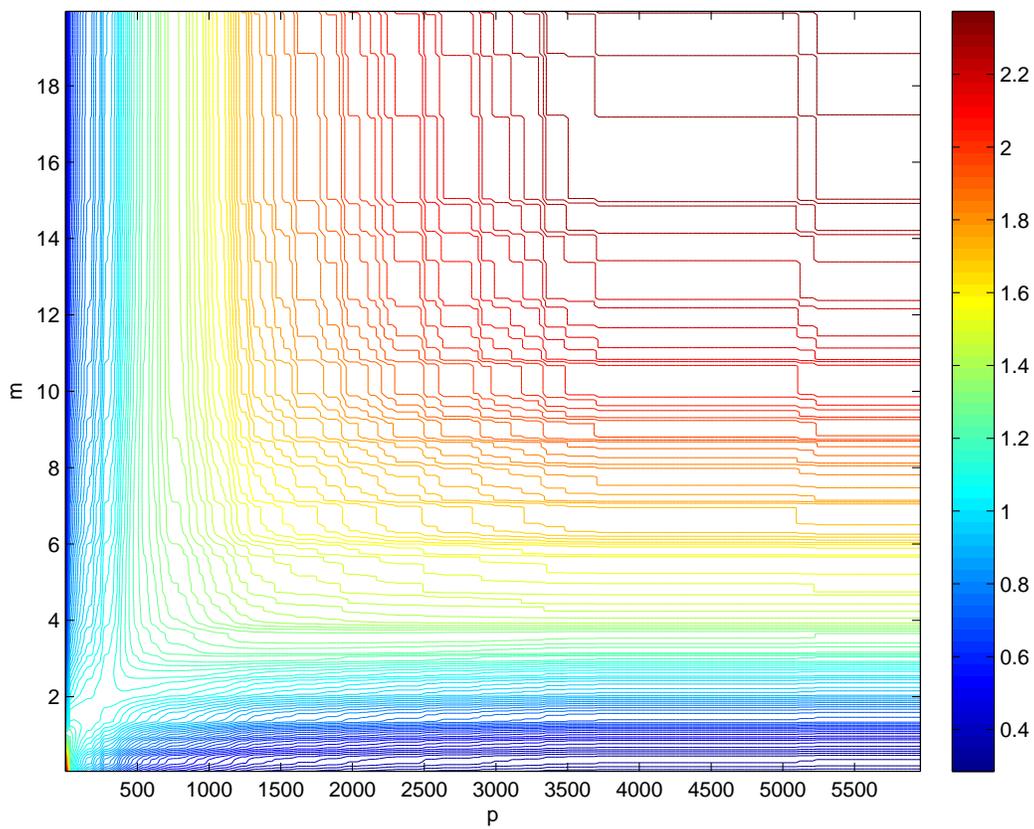}
 \caption{The color contour of
the Coupling Factor of the guess-mass model.}
\end{figure}


\clearpage
\begin{figure}
\epsscale{1.0}
\plotone{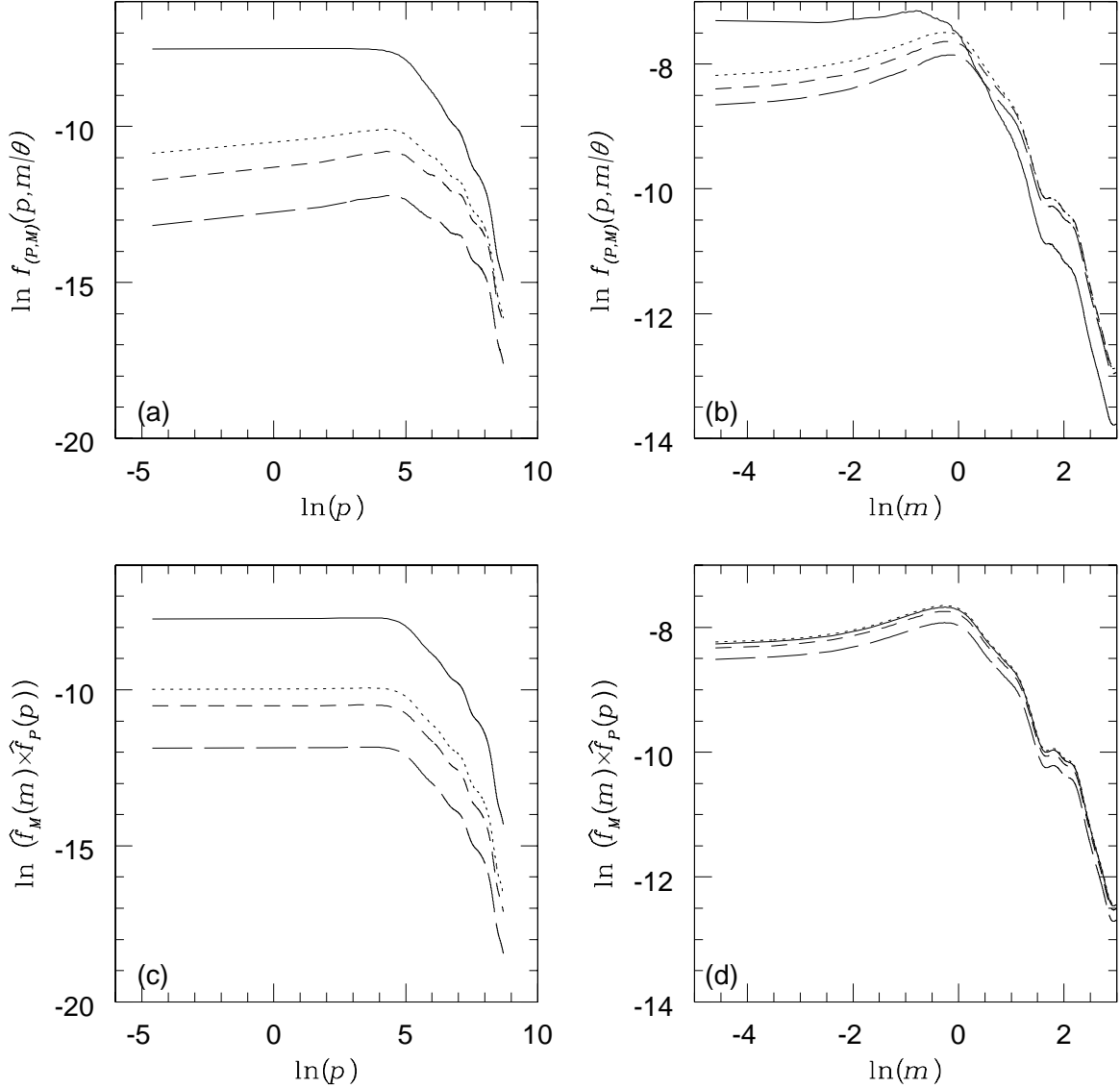}
 \caption{The period and mass
functions in logarithmic space. (a) The period functions of $m=1
M_J$ (solid curve), $m=5 M_J$ (dotted curve), $m=10 M_J$ (short
dashed curve), and $m=15 M_J$ (long dashed curve). (b) The mass
functions of $p=1$ day (solid curve), $p=50$ days (dotted curve),
$p=100$ days (short dashed curve), and $p=150$ days (long dashed
curve). (c) The independent period functions of $m=1 M_J$ (solid
curve), $m=5 M_J$ (dotted curve), $m=10 M_J$ (short dashed curve),
and $m=15 M_J$ (long dashed curve). (d) The independent mass
functions of $p=1$ day (solid curve), $p=50$ days (dotted curve),
$p=100$ days (short dashed curve), and $p=150$ days (long dashed
curve). }
\end{figure}

\end{document}